%% file: main.tex
\documentclass[twocolumn]{aastex631}
\usepackage{CJKutf8}

\shorttitle{Bright Satellites of MW analogs}
\shortauthors{Pan et al.}

\begin{document}

\title{The Merian Survey: A Statistical Census of Bright Satellites of Milky Way Analogs}

\author[0000-0002-7922-9726]{Yue Pan
\begin{CJK}{UTF8}{gbsn}(潘越)\end{CJK}}
\affiliation{Department of Astrophysical Sciences, Princeton University, 4 Ivy Lane, Princeton, NJ 08544, USA}

\author[0000-0002-1841-2252]{Shany Danieli}
\affiliation{School of Physics and Astronomy, Tel Aviv University, Tel Aviv 69978, Israel}
\affiliation{Department of Astrophysical Sciences, Princeton University, 4 Ivy Lane, Princeton, NJ 08544, USA}

\author[0000-0002-5612-3427]{Jenny E. Greene}
\affiliation{Department of Astrophysical Sciences, Princeton University, 4 Ivy Lane, Princeton, NJ 08544, USA}

\author[0000-0001-9592-4190]{Jiaxuan Li \begin{CJK}{UTF8}{gbsn}(李嘉轩)\end{CJK}}
\affiliation{Department of Astrophysical Sciences, Princeton University, 4 Ivy Lane, Princeton, NJ 08544, USA}

\author[0000-0002-3677-3617]{Alexie Leauthaud}
\affiliation{Department of Astronomy and Astrophysics, University of California, Santa Cruz, 1156 High Street, Santa Cruz, CA 95064 USA}

\author[0000-0002-0332-177X]{Erin Kado-Fong}
\affiliation{Physics Department, Yale Center for Astronomy \& Astrophysics, PO Box 208120, New Haven, CT 06520, USA}

\author[0000-0001-7729-6629]{Yifei Luo}
\affiliation{Lawrence Berkeley National Laboratory, 1 Cyclotron Road, Berkeley, CA 94720, USA}

\author[0000-0002-9816-9300]{Abby Mintz}
\affiliation{Department of Astrophysical Sciences, Princeton University, 4 Ivy Lane, Princeton, NJ 08544, USA}

\author[0000-0002-0372-3736]{Alyson Brooks}
\affiliation{Department of Physics and Astronomy, Rutgers, The State University of New Jersey, 136 Frelinghuysen Rd, Piscataway, NJ 08854, USA}

\author[0000-0003-1385-7591]{Song Huang}
\affiliation{Department of Astronomy, Tsinghua University, Beijing 100084, China}

\author[0000-0002-8040-6785]{Annika H. G. Peter}
\affiliation{Department of Physics, The Ohio State University, Columbus, OH 43210, USA}
\affiliation{Department of Astronomy, The Ohio State University, Columbus, OH 43210, USA}
\affiliation{Center for Cosmology and Astro-Particle Physics, The Ohio State University, Columbus, OH 43210, USA}

\author[0000-0001-6442-5786]{Joy Bhattacharyya}
\affiliation{Department of Astronomy, The Ohio State University, Columbus, OH 43210, USA}
\affiliation{Center for Cosmology and Astro-Particle Physics, The Ohio State University, Columbus, OH 43210, USA}

\author[0000-0001-9395-4759]{Lee S Kelvin}
\affiliation{Department of Astrophysical Sciences, Princeton University, 4 Ivy Lane, Princeton, NJ 08544, USA}

\begin{abstract}
We present a statistical census of bright, star-forming satellite galaxies around Milky Way (MW) analogs using the first data release of the Merian Survey. Our sample consists of 393 MW analogs with stellar masses $10^{10.5} < M_{\star, \rm host} < 10^{10.9} M_\odot$ at redshifts $0.07 < z < 0.09$, all central galaxies of their own dark matter halos. Using photometric selection -- including magnitude, color, angular size, photometric redshift, and size-mass cuts -- we identify 793 satellite candidates around these 393 hosts. Our selection leverages two medium-band filters targeting H$\alpha$ and [O \textsc{iii}] emission, enabling a nearly complete sample of star-forming, Magellanic Clouds-like satellites with $M_{\star, \rm sat} \gtrsim 10^{8} M_\odot$. We find that $\sim80\%$ of hosts have 0–3 massive satellites, and $13\pm4\%$ have two satellites (similar to the MW). Satellite abundance correlates with total stellar mass, and we provide significantly improved statistics for the most massive satellites at $\log_{10}[M_{\star, \rm sat}/M_{\odot}] \gtrsim 10$. The completeness-corrected radial distribution is less centrally concentrated than an NFW profile. In contrast, the Milky Way satellites are more centrally concentrated than the 50\% richest Merian systems, but are broadly consistent with the 50\% most centrally concentrated Merian systems. Our results highlight the power of medium-band photometry for satellite identification and provide a key benchmark for studying satellite quenching, environmental effects, and hierarchical galaxy formation.

\end{abstract}


\section{Introduction} \label{sec:intro}
\input{sections/1.intro}

%---------- SECTION2: Sample selection
\section{Merian Survey and Sample selection} \label{sec:sample}
\input{sections/2.survey_sel}

%---------- SECTION 3: Stats of the sample
\section{Statistics of the sample} \label{sec:stats}
\input{sections/3.stats}

\begin{figure}
    \centering
    \includegraphics[width=\columnwidth]{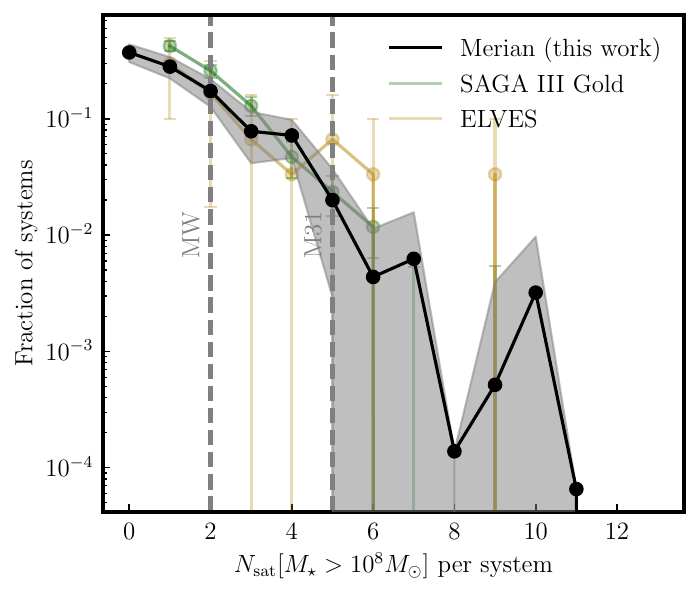}
    \caption{Fraction of systems that have satellites with $M_\star > 10^{8} M_\odot$ as a function of $N_{\rm sat}$ compared to SAGA III Gold and ELVES. MW and M31 have 2 and 5 such satellites, respectively. The shaded area is 16th and 84th percentiles from Monte Carlo simulation, computed as a combination of Poisson errors and background estimate uncertainties, the same way as Figure~\ref{fig:Nsat_dist}. Note that we apply a correction to the Merian sample corresponding to an average quenched fraction of $f_{\rm quenched} = 30\%$ due to Merian missing quenched satellites. }
    \label{fig:frac_sat_compare}
\end{figure}

%---------- SECTION 4: discussion 
\section{Discussion} \label{sec:discussion}
\input{sections/4.discussion}

%---------- SECTION 5: summary 
\section{Summary} \label{sec:summary}
\input{sections/5.summary}

\begin{acknowledgments}
This research benefited from the Dwarf Galaxies, Star Clusters, and Streams Workshop hosted by the Kavli Institute for Cosmological Physics.

% HSC-SSP and HSC-LA
The Hyper Suprime-Cam (HSC) collaboration includes the astronomical communities of Japan and Taiwan, and Princeton University. The HSC instrumentation and software were developed by the National Astronomical Observatory of Japan (NAOJ), the Kavli Institute for the Physics and Mathematics of the Universe (Kavli IPMU), the University of Tokyo, the High Energy Accelerator Research Organization (KEK), the Academia Sinica Institute for Astronomy and Astrophysics in Taiwan (ASIAA), and Princeton University. Funding was contributed by the FIRST program from the Japanese Cabinet Office, the Ministry of Education, Culture, Sports, Science and Technology (MEXT), the Japan Society for the Promotion of Science (JSPS), Japan Science and Technology Agency (JST), the Toray Science Foundation, NAOJ, Kavli IPMU, KEK, ASIAA, and Princeton University.

This paper makes use of software developed for the Vera C. Rubin Observatory. We thank the observatory for making their code available as free software at http://dm.lsst.org.

% Princeton clusters
The authors are pleased to acknowledge that the work reported in this paper was substantially performed using the Princeton Research Computing resources at Princeton University, a consortium of groups led by the Princeton Institute for Computational Science and Engineering (PICSciE) and the Office of Information Technology's Research Computing.

\end{acknowledgments}

\software{astropy \citep{2013A&A...558A..33A,2018AJ....156..123A}, Source Extractor \citep{1996A&AS..117..393B}}

%----------Appendix 
\appendix
\twocolumngrid

\section{Halo group mass and total stellar mass}\label{sec:appendix1}
\begin{figure}[htbp]
    \centering
    \includegraphics[width=\columnwidth]{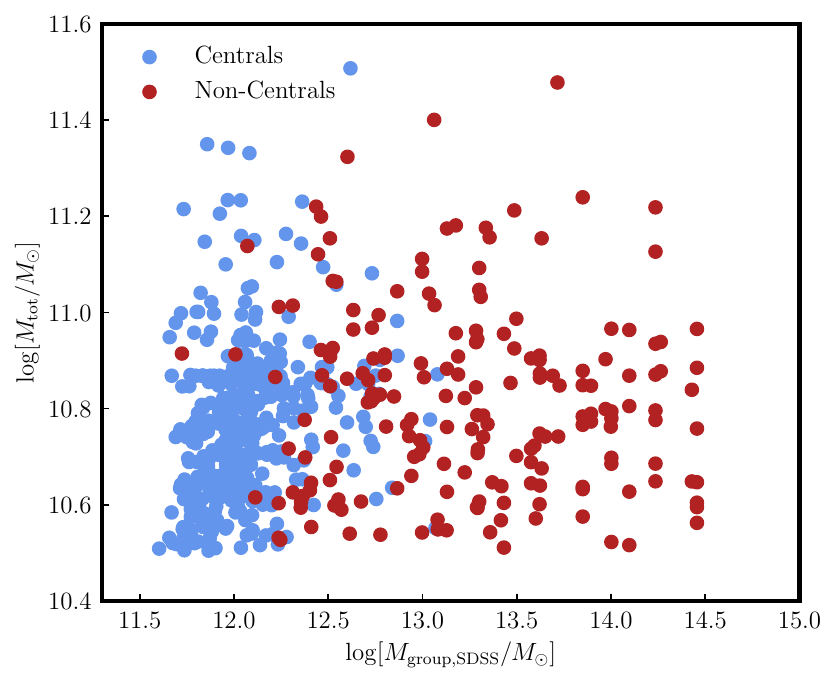}
    \caption{$M_{\rm tot}$ as a function of $M_{\rm group}$ from SDSS DR13 for all hosts in our sample that have a SDSS group mass. Centrals and non-centrals are identified and labeled in SDSS. Galaxies in more massive groups in SDSS tend to be non-centrals. For central hosts, we find a moderate positive monotonic correlation between group mass and total stellar mass (Spearman $\rho_s=0.309$, 95\% CI $[0.209,\,0.403]$, $p=3.85\times10^{-10}$, $N=393$)}. We take only centrals in SDSS DR13 as our final cut for the host selection.
    \label{fig:Mtot_Mgroup}
\end{figure}

In Figure~\ref{fig:Mtot_Mgroup}, we show $M_{\rm tot}$ as a function of $M_{\rm group}$ from SDSS DR13 for all 393 hosts in our sample that have an SDSS group mass. We separate the hosts using the \texttt{is\_central} label in the SDSS group catalog. Non-centrals tend to reside in massive groups with $\log[M_{\rm tot/M_\odot}] \gtrsim 13$ as they could potentially contain multiple primaries, while centrals tend to reside in smaller group with $\log[M_{\rm tot/M_\odot}] \lesssim 13$. For central hosts, we find a moderate positive monotonic correlation between group mass and total stellar mass (Spearman $\rho_s=0.309$, 95\% CI $[0.209,\,0.403]$, $p=3.85\times10^{-10}$, $N=393$).

\section{Bayesian Estimation of Background-Subtracted Satellite Abundance}\label{sec:appendixb}

To estimate the true satellite abundance $N_{\rm sat}$ while accounting for Poisson noise and background contamination, we adopt a Bayesian framework with a non-informative Jeffreys' prior. Let $k_{\rm obs}$ be the observed number of candidate satellites and $b$ be the expected background contamination. We model the observed counts as a Poisson process:
\[
P(k_{\rm obs} \mid \lambda) = \frac{(\lambda + b)^{k_{\rm obs}}}{k_{\rm obs}!} e^{-(\lambda + b)},
\]
where $\lambda$ is the true signal count (i.e., the background-subtracted satellite abundance). 

To construct the posterior, we adopt a Jeffreys' prior for the Poisson rate parameter:
\[
P(\lambda) \propto \lambda^{-1/2},
\]
which reflects maximal ignorance about the scale of $\lambda$ and is known to perform well in low-count regimes.

The resulting posterior distribution for $\lambda$ is a Gamma distribution:
\[
\lambda \sim \mathrm{Gamma}(k_{\rm obs} - b + 0.5,\, 1),
\]
where the first parameter is the shape and the second is the rate. The posterior mean, which we adopt as our estimate of the true satellite abundance, is:
\[
\mathbb{E}[\lambda] = k_{\rm obs} - b + 0.5.
\]

This estimate is well-defined even when $k_{\rm obs} \approx b$ or zero, providing a smooth and statistically principled correction for background contamination. This method is used to generate the background-subtracted counts shown as the solid black line in Figure~\ref{fig:Nsat_dist}.

\bibliography{references}{}
\bibliographystyle{aasjournal}

\end{document}

%% file: sections/1.intro.tex
The study of satellite galaxies offers a crucial window into the mechanisms that shape their formation and evolution. Satellites are not merely passive tracers of their host galaxy’s gravitational potential, but play an active role in shaping their hosts through interactions such as tidal stripping, mergers, and environmental quenching \citep[e.g.,][]{Tollerud.etal.2011, Wetzel.etal.2013, Fillingham.etal.2015, Saeedzadeh.etal.2023, Joshi.etal.2025}. The abundance, distribution, and properties of satellite galaxies provide key constraints on the hierarchical formation of structure in a $\Lambda$CDM universe, offering insight into how host galaxies evolve over time \citep[e.g.,][]{Wechsler.Tinker.2018}. However, a comprehensive understanding of satellite populations requires statistical samples that span a wide range of environments, masses, and evolutionary states.

The characterization of satellites around individual hosts  -- and, in particular, those of the Milky Way -- have played a key role in driving our understanding of the physical processes that shape the evolution of satellite galaxies at large. However, whether the Milky Way’s satellite configuration is representative of the broader population of similar-mass galaxies remains an open question \citep[e.g.,][]{Tollerud.etal.2011, Boylan-Kolchin.etal.2012, Samuel.etal.2020}. The Milky Way hosts a diverse array of satellites, including the Large and Small Magellanic Clouds (LMC and SMC), which are among the most massive satellites observed around a Milky Way-like system. These satellites are actively interacting with the Milky Way, with recent studies suggesting that they may significantly impact the Galactic disk, halo, and overall mass distribution \citep[e.g.,][]{Besla.etal.2012, Erkal.etal.2019, Patel.etal.2020}. The satellite system of the Milky Way also provides a critical testing ground for dark matter models, not only in terms of abundance but also in the detailed properties of satellite galaxies. Discrepancies between the observed and predicted quenched fractions at small host radii \citep[e.g.,][]{Font.etal.2022, Samuel.etal.2022, Pan.etal.2023, Li.etal.2025b}, as well as the radial distribution of satellites \citep[e.g.,][]{2018PhRvL.121u1302K,2021arXiv210609050K,Samuel.etal.2020, Carlsten.etal.2020}, suggest that baryonic physics, environmental effects, and artificial disruption of satellites in simulations may be more complex than commonly captured in standard semi-analytic or hydrodynamical models. These fine-grained tensions – the overabundance of quenched satellites at small radii in simulations or the steep radial profiles – highlight potential challenges to $\Lambda$CDM and/or galaxy formation on small scales and point toward the need for improved modeling of satellite quenching mechanisms and infall histories.

Targeted studies of nearby Milky Way analogs have provided valuable insight into the diversity of satellite systems in different environments, either using direct distances \citep[e.g., ][]{Crnojevic.etal.2016, Muller.etal.2019,Toloba.etal.2016, Mutlu-Pakdil.etal.2022, Mutlu-Pakdil.etal.2024} or with statistical background subtraction \citep{Tanaka.etal.2018,Nashimoto.etal.2022}.

Larger-volume efforts have systematically explored satellite populations complete to classical dwarf masses across dozens to hundreds of Milky Way analogs. The Satellites Around Galactic Analogs (SAGA) Survey \citep{Geha.etal.2017, Mao.etal.2021, Mao.etal.2024, Geha.etal.2024, Wang.etal.2024} focuses on 101 Milky Way–mass galaxies at distances of 25--40.75~Mpc, selected by $K$-band luminosity and isolation. The survey includes 378 spectroscopically confirmed satellites and provides strong constraints on satellite abundance and star formation activity. The Exploration of Local VolumE Satellites (ELVES) survey \citep{Carlsten.etal.2021b, Carlsten.etal.2022} complements SAGA by targeting galaxies within $d<12$~Mpc using deep imaging. ELVES detects satellites down to $M_{\star, \rm sat} \sim 10^{5.5} M_{\odot}$. The hosts, spanning $10^{10} - 10^{11} M_{\odot}$, are surveyed to at least 150 kpc, with most covered to 300 kpc. The final dataset comprises 338 confirmed satellites and 106 candidates.

For yet larger volumes, it has thus far only been possible to study LMC-mass satellite systems with statistical approaches, typically across tens of thousands of host galaxies \citep{Sales.etal.2013, Wang.etal.2021, Wang.etal.2025}. 
At higher redshifts, similar statistical techniques have been applied to measure the satellite luminosity function and spatial distributions around massive hosts, revealing how satellite abundance and radial profiles evolve with cosmic time \citep[e.g.,][]{Nierenberg.etal.2012, Nierenberg.etal.2016}.

Building on these previous efforts, the Merian Survey provides a complementary strategy for studying satellite galaxies by targeting star-forming dwarf satellites using deep, wide-field optical imaging \citep{Luo.etal.2023, Danieli.etal.2024}. Specifically, Merian is optimized to study the properties of bright star-forming dwarf galaxies over large volumes \citep[e.g.,][]{Mintz.etal.2024}. Merian uses DECam medium-band imaging to isolate H$\alpha$ and [O~\textsc{iii}] emission, enabling the identification of star-forming satellites around $\sim$400 Milky Way analogs at $0.07 < z < 0.09$ from imaging alone--without the need for targeted spectroscopy (as in SAGA)--and over larger volumes than ELVES. Compared to the previous studies, Merian substantially increases the number of hosts while maintaining a consistent satellite mass range and accurate distance measurements via photometric redshift (photo-$z$), allowing for a more robust characterization of satellite populations. Merian’s photometric selection approach provides an avenue to study low-surface-brightness satellites and diffuse star-forming galaxies that may be missed in previous spectroscopic surveys. This study provides a useful observational constraint for theoretical models of satellite abundance, quenching, and hierarchical galaxy formation \citep[e.g.,][]{Fakhouri.etal.2010, Rodriguez-Gomez.etal.2016, Behroozi.etal.2019, Christensen.etal.2024}.

The paper is structured as follows. In Section~\ref{sec:survey_summary}, we provide an overview of the Merian Survey, describing its observational strategy and dataset. Section~\ref{sec:host_select} outlines the selection criteria for Milky Way analog host galaxies, while Section~\ref{sec:sat_select} details the methodology for identifying satellite galaxies, including photometric selection criteria and contamination corrections. Section~\ref{sec:stats} presents our key statistical results on satellite systems. 
Finally, Section~\ref{sec:comparison_SAGA_ELVES} compares the Merian satellite population to previous surveys, and Section~\ref{sec:LMC_SMC_context} places the Milky Way’s satellite system in a cosmological context. In Section~\ref{sec:summary}, we summarize our conclusions and discuss implications for future studies of satellite galaxies.

Throughout this paper, we adopt a flat $\Lambda$CDM model with $H_0=70$ km s$^{-1}$ Mpc$^{-1}$ and $\Omega_m=0.3$, and a Salpeter initial mass function (IMF) \citep{Salpeter.1955}.

%% file: sections/2.survey_sel.tex
%---------- SUBSECTION 2.1: Overview of the Merian Survey
\subsection{Overview of the Merian Survey} \label{sec:survey_summary}

The Merian Survey is a wide-field optical imaging survey optimized to study the properties of bright star-forming dwarf galaxies over large volumes \citep{Luo.etal.2023, Danieli.etal.2024}. The survey employs two custom-designed medium-band filters, N540 ($\lambda_c = 5400$\AA, $\Delta \lambda = 210$\AA) and N708 \citep[$\lambda_c = 7080$\AA, $\Delta \lambda = 275$\AA;][]{Luo.etal.2023}, specifically built for the Dark Energy Camera (DECam) on the 4-m Blanco telescope. These filters are tailored to detect H\(\alpha\) and [O \textsc{iii}]$\lambda 5007\rm\AA$ line emission at redshifts $0.06\lesssim z\lesssim 0.1$, 
allowing for accurate photometric redshifts ($\sigma_{\Delta z/(1+z)} \approx 0.01$) in this redshift window (Luo et al., in prep).

Covering an area of approximately 750 square degrees, the Merian Survey was designed to overlap with the wide layer of the Hyper Suprime-Cam Subaru Strategic Program (HSC-SSP) equatorial fields \citep{Aihara.etal.2018a, Aihara.etal.2022}. The survey design principles were guided by the need for accurate photometric redshifts (photo-$z$), high sample completeness, and optimized galaxy-galaxy lensing signal-to-noise for dwarf galaxies \citep{Leauthaud.etal.2020, Thornton.etal.2024, To.etal.2025, Treiber.etal.2025}. By leveraging the publicly available HSC-SSP dataset, which offers deep and high spatial resolution wide-field imaging in five broad-band filters ($grizy$), Merian is capable of exploring a rich parameter space that includes stellar masses, sizes, star formation rates, and metallicities down to \(M_\star \approx 10^8 \,M_\odot\) \citep{Danieli.etal.2024}.

%----------FIG 1: sample systems with 0, 1, 2, 3 satellites
\begin{figure*}
    \centering
    \includegraphics[width=\textwidth]{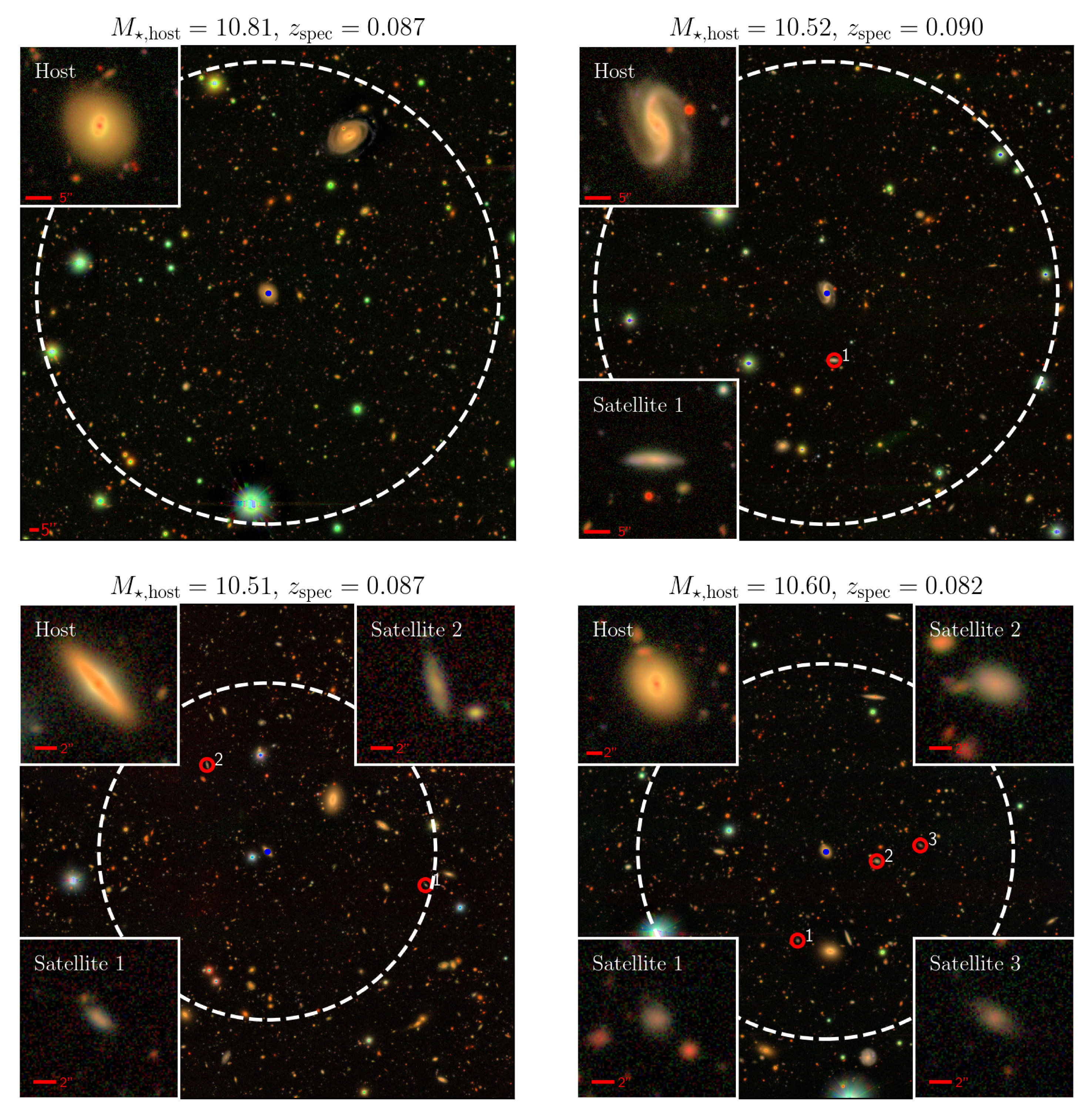}
    \caption{Color stamps from HSC are displayed for four representative systems in our sample, showcasing cases with 0, 1, 2, and 3 satellite candidates, arranged from the top left to the bottom right, respectively, together with the log stellar mass and spectroscopic redshift. Each panel features the main figure, which illustrates the host environment extending out to its virial radius, represented by the white dashed circle. To provide more detail, a zoomed-in view of the host galaxy is shown in the upper left corner of each panel. For systems with satellites, additional insets highlight the satellites, emphasizing their positions and characteristics within the host environment.
}
    \label{fig:figure1}
\end{figure*}

In the Merian Data Release 1 \citep[DR1;][]{Danieli.etal.2024}, we utilize the Rubin Observatory LSST Science Pipelines\footnote{https://pipelines.lsst.io/} \citep{Bosch.etal.2018, Bosch.etal.2019} to perform joint photometry across the two DECam medium-band images and five HSC broad-band $(grizy)$ images \citep{Aihara.etal.2018a, Aihara.etal.2018b, Aihara.etal.2019, Aihara.etal.2022}. DR1 covers a total full color-full depth area of $234\,\mathrm{deg}^2$. To ensure accurate color measurements that are essential for photometric redshifts, we employ the Gaussian-Aperture-and-PSF (GAaP) photometry technique\footnote{https://dmtn-190.lsst.io/} \citep{Kuijken.etal.2008}. GAaP is optimized for consistent PSF- and aperture-matched photometry across different seeing conditions and telescopes, making it ideal for our multi-system survey. This approach mitigates systematic uncertainties that would arise from aperture photometry, ensuring robust results despite varying image qualities. \texttt{CModel} photometry  \citep{Bosch.etal.2018} is also included in our DR1 photometric catalog.

To accommodate sources from varying depths and filter combinations, we generate a Hierarchical Equal Area isoLatitude Pixelization (\texttt{HEALPix}; \citealt{Gorski.etal.2005}) projection map for Merian DR1. The \texttt{HEALPix} mask identifies regions with data in all seven bands ($grizyN540N708$; full color) and with effective exposure times $t_{\rm eff} > 1200\,\mathrm{sec}$ for $N708$ and $t_{\rm eff} > 1800\,\mathrm{sec}$ for $N540$ (full depth). We define the effective exposure time as $t_{\rm eff} = \tau t_{\rm exp}$, with $t_{\rm exp}$ representing the open shutter time and $\tau$ accounting for factors such as atmospheric transmission, sky brightness, and seeing \citep[for a full characterization of effective exposure time in the context of the Merian Survey, see][]{Danieli.etal.2024}. The \texttt{HEALPix} mask has a resolution of approximately $50\arcsec$, which is much smaller than the Merian DR1 full-color-full-depth field of view of 234 deg$^2$. We use this \texttt{HEALPix} map to determine the fraction of each host galaxy's virial area (Section~\ref{sec:mag_color_size_cut}) that is masked due to artifacts.

While the Merian Survey covers a redshift range of \(0.06 < z < 0.1\), our analysis is specifically focused on the narrower range of \(0.07 < z < 0.09\). This restriction minimizes instrumental effects, such as variations of the filter transmission curve across the focal plane, that are more pronounced at the edges of the broader redshift range, potentially compromising the accuracy of the photometric redshift measurements. By concentrating on this central redshift interval, we aim to ensure higher selection purity and better characterization of the targeted dwarf galaxies.

We use every source in the Merian DR1 photometric catalog where:
\begin{enumerate}
    \item The source is not masked by the HSC-SSP PDR3 bright star mask.
    \item The source is in the Merian full-depth full-color area.
    \item The source meets various photometry quality criteria (see details in \citealt{Danieli.etal.2024})
    \item The source is not a star according to its $i$-band extendedness value used for star-galaxy separation.
\end{enumerate}

%---------- SUBSECTION 2.2: Host galaxy selection
\subsection{Host galaxy selection}\label{sec:host_select}

Our Milky Way analog hosts are selected from a cross-matched catalog of sources with spectroscopic redshifts in the range $0.07 < z < 0.09$, as detailed in \citet{Mintz.etal.2024}. The host catalog is constructed by combining the Merian DR1 catalog with spectroscopic data from the Galaxy and Mass Assembly survey \citep[GAMA;][]{Baldry.etal.2018}, the Sloan Digital Sky Survey \citep[SDSS;][]{Almeida.etal.2023}, and proprietary Merian spectroscopic observations (Luo et al. 2025, in preparation). The Merian footprint overlaps with GAMA fields G02, G09, G12, and G15, and we perform a cross-match using a $0.3\arcsec$ search radius.

Stellar masses of the Merian host galaxies are derived using the $g-r$ color-based stellar mass-to-light ratio relation from \citet{Bell.etal.2003}:

\begin{equation}\label{eqn:mass_color}
\log_{10}(M_\star/L_{r}) = -0.306 + 1.097 (g-r)
\end{equation}
This empirical relation allows us to estimate the stellar mass of each galaxy directly from its observed $g-r$ color and $r$-band luminosity. The use of this method ensures consistency with prior studies of Milky Way analogs \citep{Carlsten.etal.2021, Mao.etal.2021, Mao.etal.2024, de.los.Reyes.etal.2024}, while also providing a robust estimate of stellar mass that accounts for variations in stellar populations and dust attenuation. We do not apply $k$-corrections to the magnitudes, as it is expected to be negligible within our redshift range and the large stellar mass range of the host galaxies.

We note several caveats to this color–$M_\star/L$ estimator. 
While these effects are well understood and stellar masses derived from colors 
tend to be reasonably robust overall, some systematic uncertainties remain. 
The calibration depends on the adopted stellar population synthesis models, 
dust law, and IMF, such that alternative choices can shift the zero point by 
$\sim$0.1--0.2\,dex. A single color cannot fully break age--metallicity--dust 
degeneracies: dust reddening can mimic older populations, and recent star 
formation bursts can temporarily ``outshine'' older stellar populations, 
especially at low mass or high sSFR. The calibration is performed in the 
rest frame, so neglecting $k$-corrections can introduce small redshift-dependent 
biases, with additional scatter arising from $g/r$ aperture differences and 
color gradients. These limitations reflect the broader, well-known systematic 
uncertainties inherent to SED modeling \citep[e.g.][]{Pacifici.etal.2023,Iyer.etal.2025}.

Milky Way analogs are identified within this sample using a selection criterion based on spectroscopic redshift, stellar mass, and group central flags constructed by applying the group finder in \citet{Lim.etal.2017} to Sloan Digital Sky Survey Data Release 13 \citep[SDSS DR13;][]{Albareti.etal.2017}:  
\begin{enumerate}
    \item $10.5 < \log[M_{\star, \rm host}/ M_\odot] < 10.9$
    \item $0.07 < z < 0.09$
    \item \texttt{is\_central} in SDSS DR13 using the group finder in \citet{Lim.etal.2017}\footnote{\url{https://gax.sjtu.edu.cn/data/Group.html}}
\end{enumerate}

This mass range represents a 0.4 dex bin centered around the widely accepted value for the Milky Way's stellar mass, $\log (M_\star / M_\odot) = 10.7$, as reported by \citet{Licquia.Newman.2015}. The lower bound of $\log(M_\star/M_\odot) = 10.5$ ensures that we include galaxies with stellar masses comparable to the Milky Way, which are expected to reside in similar halo mass regimes. Meanwhile, we apply the upper stellar mass cut of $\log(M_\star/M_\odot) = 10.9$ to exclude high halo masses. Including higher-mass systems risks misidentifying our selected hosts as satellites of brighter galaxies in such groups, thereby complicating the comparison to the Milky Way's relatively isolated environment. To avoid contamination from groups, we further exclude massive group environments using a group catalog, ensuring that our Milky Way analogs more closely resemble the Milky Way in both stellar mass and environment (see below). Our systems still span a wide range of halo masses, and we explore the dependence of satellite properties on this range below.

To ensure that all hosts in our sample are truly central galaxies of their respective halos, we impose the \texttt{is\_central} selection criterion in SDSS DR13, which is identified using the group finder in \citet{Lim.etal.2017}. This prevents contamination from massive satellite galaxies that might otherwise be misclassified as Milky Way analogs. As shown in Figure~\ref{fig:Mtot_Mgroup} in the appendix, galaxies in more massive groups are predominantly classified as non-centrals. Our application of the \texttt{is\_central} flag removes these systems, maintaining a more direct comparison to the Milky Way, which is the central galaxy of its own halo. After these cuts, we obtain a total of 393 Milky Way analogs in our host galaxy sample.

%----------table1
\begin{deluxetable}{lc}
\tablecaption{Downselection from the full satellite photometric catalog to the satellite candidate sample.\label{tab:downselection}}
\tablehead{
\colhead{Selection step} & \colhead{Objects remaining}
}
\startdata
Photometric cuts (magnitude, size, color)                 &   41,000 \\
Integrated $p(z)$      &  1191 \\
Size--mass                   &   972 \\
Duplicates            &    950 \\
Deblending                   &    793 \\
Background subtraction                   &    451 $\pm$47 \\
\enddata
\tablecomments{Before background subtraction, our sample consists of 793 satellite candidates. Since the total background number for all 393 hosts is 342$\pm$38, the final sample has 451$\pm$47 satellite candidates (Section~\ref{sec:bkg_sub})}.
\end{deluxetable}

%---------- SUBSECTION 2.2: Satellite galaxy selection
\subsection{Satellite galaxy selection} \label{sec:sat_select}
%---------- SUBSECTION 2.2.1: Magnitude, color, size cut
\subsubsection{Magnitude, Color, and Size Cuts} \label{sec:mag_color_size_cut}

To identify satellite candidates around each host galaxy, we first estimate the host halo mass using the stellar mass--halo mass relation from \citet{Behroozi.etal.2010}. We then use \texttt{Halotools}\footnote{\url{https://github.com/astropy/halotools}} \citep{Hearin.etal.2017} to compute the corresponding virial radius, $R_{\rm vir}$, adopting the \texttt{"vir"} mass definition. This definition corresponds to a spherical overdensity criterion in which the mean density within the virial radius equals a redshift-dependent overdensity factor, $\Delta_{\mathrm{vir}}(z)$, times the mean matter density of the universe, following the spherical collapse model. The mean $R_{\rm vir}$ for our host sample is 330 kpc, the smallest is 256 kpc, and the largest is 446 kpc. To identify objects that are physically associated with each host galaxy, we search for objects in the Merian DR1 catalog that fall within the projected virial radius (\( R_{\rm vir} \)) of the nearest host on the sky. We assume that each source is at the same redshift as its corresponding host and calculate its projected separation accordingly.

To reduce contamination from the central regions of the massive host galaxy, where stellar crowding and instrumental effects such as background oversubtraction can obscure faint sources, we exclude all sources within $0.1\, R_{\rm vir}$ of the hosts ($\sim 30$ kpc, on average). This radial exclusion helps ensure a cleaner sample of candidate satellites by minimizing contamination from the host's light profile and unresolved background sources.

Some regions within the virial radius may be masked due to bright stars, diffraction spikes, or other artifacts in the HSC imaging footprint. To address this, we employ the \texttt{HEALPix} mask to estimate the fraction of the virial area that is effectively masked. If more than 10\% of a host's virial region (excluding the inner $0.1 R_{\rm vir}$) is masked, we exclude that host from our final sample. Less than 5\% of the systems are excluded this way. This masking criterion ensures that our measurements of satellite abundance are not biased by incomplete coverage or observational artifacts.

After selecting sources within the radial range of $0.1\, R_{\rm vir}$ to $R_{\rm vir}$, we apply three initial cuts based on the \texttt{cModel} magnitude, \texttt{GAaP} color\footnote{Here, we use the \texttt{GAaP} 1.0" photometry}, and \texttt{cModel} $r_e$ to remove faint background objects and unresolved sources:
\begin{equation}
i_{\rm cModel} < 23\,\mathrm{mag},\quad 0 < (g-r)_{\rm GAaP} < 1,\quad r_e > 0.5\arcsec
\end{equation}

The magnitude cut of \( i_{\rm cModel} < 23 \) is chosen to ensure that we include only relatively bright sources while minimizing contamination from unresolved background objects. As shown in Figure 11 in \citet{Danieli.etal.2024}, the number of detected sources in the Merian DR1 catalog exhibits a steep decline at \( m_{N708} \sim 23 \) mag, indicating a drop in sensitivity beyond this threshold. By adopting this magnitude limit, we mitigate the risk of magnitude incompleteness while maintaining a robust selection of sources with well-measured photometry.

\begin{figure}
    \centering
    \includegraphics[width=\columnwidth]{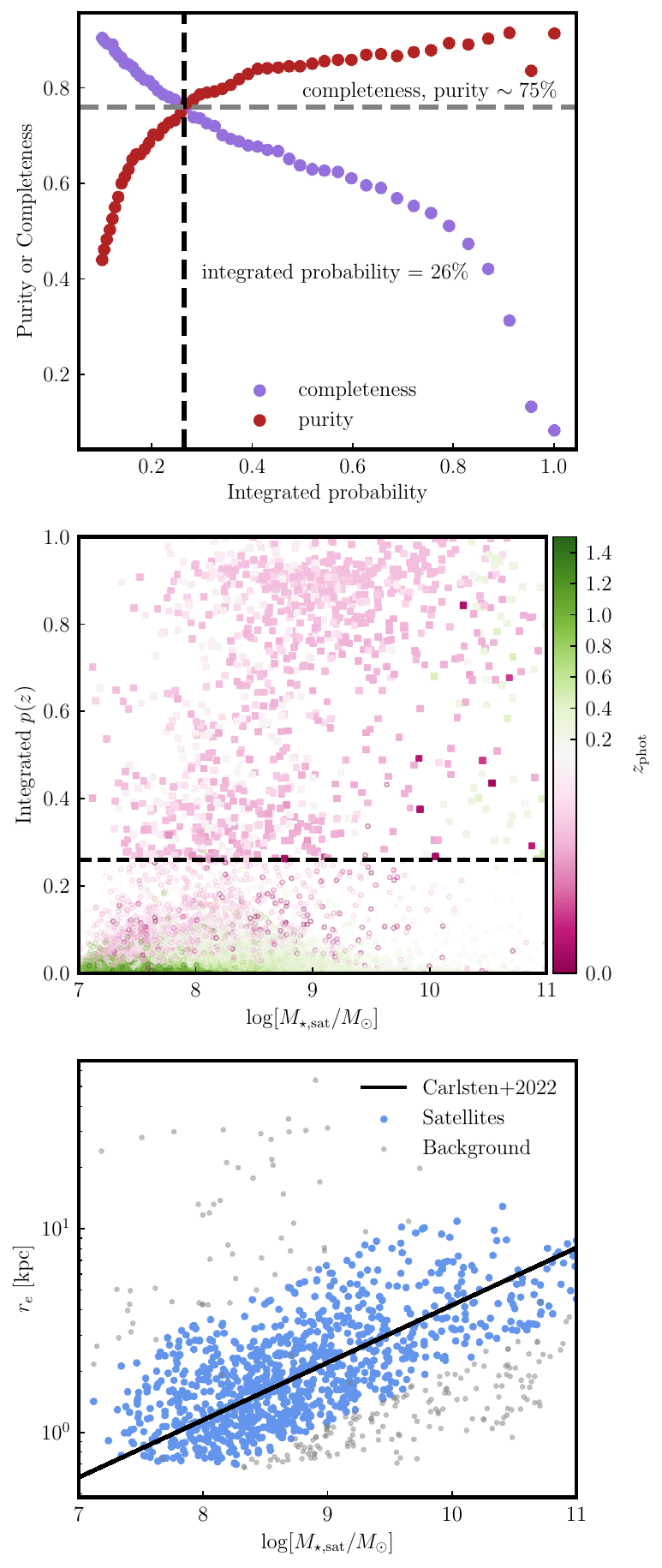}
    \caption{\emph{Top}: Purity and completeness of our COSMOS spec-$z$ sample as a function of integrated probability $p(z)$ (Section~\ref{sec:photz_cut}). As expected, completeness decreases while purity increases with increasing integrated $p(z)$. We set the threshold where they intersect, at $p(z) = 0.26$ (26\%), achieving $\sim$75\% completeness and purity.  \emph{Middle}: Integrated $p(z)$ vs. satellite $M_\star$, color-coded by $z_{\rm phot}$ from \texttt{EAZY}, for sources after the initial cuts (Section~\ref{sec:mag_color_size_cut}). Green symbols are background sources and pink are in-band sources. Of $\sim$41,000 sources, $\sim$1,200 pass the $p(z) > 0.26$ cut (squares).  \emph{Bottom}: Applying a size--mass cut based on the LTG size--mass relation \citep{Carlsten.etal.2022} to these $\sim$1,200 sources leaves $\sim$970 (blue).  
}
    \label{fig:test_pz_sizemass}
\end{figure}

The $(g-r)_{\rm GAaP}$ color range of $0 < (g-r) < 1$ selects sources with plausible colors for low-mass galaxies, effectively excluding extremely red or blue objects unlikely to be physical companions \citep{Greco.etal.2018, Carlsten.etal.2021}. Finally, the angular size cut $r_e > 0.5\arcsec$ excludes unresolved sources that might be stars or background objects at high redshift. The half-light radii $r_e$ are computed from the quadrupole moments of the shape parameters, derived from the DR1 \texttt{deepCoadd\_obj} catalog \citep{Joachimi.etal.2013}. This approach to measure size is widely used in morphological selection and weak lensing studies \citep[e.g.,][]{Chang.etal.2013}. Our approach ensures consistent and reliable size measurements across all sources in the sample. Real satellites with $\log[M_\star/M_\odot] = 8$ have a physical size of $\sim1$ kpc, adopting size--mass relation for late-type dwarf galaxies from \citet{Carlsten.etal.2021} (see more in Section~\ref{sec:size_mass_cut}, see also Figure~\ref{fig:test_pz_sizemass}). At $z = 0.08$, this corresponds to an angular size of $\sim 0.7\arcsec$, so real satellites would not be removed by this angular size cut. While this size cut is effective at removing point sources, it may also exclude compact satellite galaxies that fall below the stellar mass limit of our sample.

By combining these magnitude, color, and size cuts, we effectively filter out contaminants and retain a clean sample of candidate satellite galaxies for further analysis.

%---------- SUBSECTION 2.3.2: Photometric redshift cut
\subsubsection{Photometric Redshift Cut} \label{sec:photz_cut}

One of the primary goals of the Merian survey is to obtain precise photo-$z$ for sources in the HSC field. This is achieved by leveraging a combination of five HSC broad bands ($grizy$) and two Merian medium bands (N540 and N708), which are specifically designed to target the [O~\textsc{iii}] and H$\alpha$ emission lines, respectively. To derive photo-z values for each source in Merian DR1 (Luo et al., in prep), we utilize the template-fitting photo-z code \texttt{EAZY} \citep{Brammer.etal.2008}, applied to 7-band \texttt{GAaP} photometry measured within a 1.0$\arcsec$ aperture. We only derive photo-$z$'s for galaxies flagged as science-ready sources (\texttt{SciUse}=1) in the Merian DR1 photometric catalog \citep{Danieli.etal.2024}. The Merian photo-$z$ performance is described in Luo et al., in prep.

The \texttt{EAZY} outputs from the Merian DR1 photo-$z$ catalog include the least-$\chi^2$ photo-z value, along with five point estimates at the 2.5\%, 16\%, 50\%, 84\%, and 97.5\% confidence intervals of the photo-$z$ probability density function (PDF). Additionally, it provides a $\chi^2$ distribution of the photo-z across 184 grid points spanning the redshift range $0 < z < 1.5$. This $\chi^2$ distribution is converted into a PDF of the photo-$z$, $p(z)$, by incorporating prior information based on the $r$-band luminosity function from \citet{Davidzon.etal.2017}. 

To evaluate the likelihood of a source having a photo-$z$ within 0.06 and 0.1, we integrate the $p(z)$ over the redshift range $0.06 < z < 0.1$, yielding an integrated $p(z)$ value within this range. This value quantifies the probability of the source residing within the desired redshift range. We choose to integrate in the range $0.06-0.1$ instead of our host spec-$z$ range $0.07-0.09$ because our Merian medium band photo-$z$ can only provide information about the source being in or out of the $0.06-0.1$ redshift range without further refinement. We use integrated $p(z)$ instead of a traditional photo-$z$ point estimation like \texttt{z500}, since integrated $p(z)$  gives a more complete and pure sample, plus being more flexible to apply a cut with.

\subsubsection{Determining $p(z)$ Threshold for Sample Selection}

To select in-band sources, we need an appropriate threshold for the integrated $p(z)$ value. We use a spectroscopic redshift (spec-z) sample from low-mass galaxies in the COSMOS field to select in-band sources, which enables us to examine how completeness and purity change as a function of different integrated $p(z)$ thresholds. Our spec-z sample comprises galaxies with \( 7.5 < \log(M_\star/M_\odot) < 9.5 \) and \( z < 0.25 \), selected from the COSMOS2015 catalog \citep{Laigle.etal.2016}. It also includes candidates for Ly$\alpha$ emitters at \( z \approx 3.4 \) and \( z \approx 4.8 \), identified through flux excesses in the N540 and N708 filters, respectively. 

The sample consists of 3,914 dwarf galaxies within these mass and redshift ranges, observed using IMACS, DEIMOS, and DESI instruments \citep{Danieli.etal.2024}. Approximately 200 of these sources have in-band spec-z measurements ($0.06 < z < 0.1$). Further details on Merian photometric redshifts and spectroscopic data will be discussed in Luo et al. (in preparation).

We define sample completeness as:
\begin{equation}
\frac{\text{\# of spec-z in-band \& integrated } p(z) > \text{ threshold}}{\text{\# of spec-z in-band}},
\end{equation}
and sample purity as:
\begin{equation}
\frac{\text{\# of spec-z in-band \& integrated } p(z) > \text{ threshold}}{\text{\# of integrated } p(z) > \text{ threshold}}.
\end{equation}

We calculate the sample completeness and purity for a range of integrated $p(z)$ thresholds, shown in Figure~\ref{fig:test_pz_sizemass}. For a threshold value of 0.26 (26\%), the spec-$z$ sample achieves a balance of high purity and completeness, both approximately 75\%. Consequently, we adopt this value as our integrated $p(z)$ threshold for filtering out background sources (see top panel of Figure~\ref{fig:test_pz_sizemass}).

The middle panel of Figure~\ref{fig:test_pz_sizemass} shows the integrated $p(z)$ values for sources passing the initial cuts described in Section~\ref{sec:mag_color_size_cut}. In general, sources with an integrated $p(z) > 0.26$ have $z_{\rm phot}$ values (point estimate from \texttt{EAZY}) within the desired range, while those with lower integrated $p(z)$ values are typically at higher redshifts or belong to the background population. Moreover, the few interlopers with integrated $p(z) > 0.26$ but out-of-band $z_{\rm phot}$ values tend to be massive galaxies with $\log [M_{\star, \rm sat}/M_\odot] > 10$.  These results emphasize the effectiveness of the $p(z)$ threshold in isolating in-band sources while minimizing contamination from the foreground and background. For our source selection, we require integrated $p(z) > 0.26$, and a physical photometric redshift with $z_{\rm phot} > 0$.

%---------- SUBSECTION 2.2.2: Size--mass cut
\subsubsection{Size--mass cut} \label{sec:size_mass_cut}

To further exclude background sources, we apply the size--mass relation for late-type dwarf galaxies. Background sources typically appear smaller in size for a given magnitude, making them clear outliers in this relation. Using the spectroscopic redshift (spec-z) from each satellite galaxy's host, we convert its angular size in the $N708$ filter into physical size. This conversion allows us to identify and filter out sources that, if assumed to be at the distance of the host galaxy, would deviate significantly from the expected size-mass relation. Given that the Merian survey is designed to detect H$\alpha$ for photo-z accuracy, the majority of our potential satellite galaxies exhibit signs of active star formation. Consequently, we adopt the size-mass relation for late-type dwarf galaxies derived from the Exploration of Local VolumE Satellites (ELVES) Survey:

\begin{equation}
\log(r_e/\text{pc}) = a + b \log(M_\star/M_\odot),
\end{equation}
where $a = 0.805_{-0.245}^{+0.242}$, $b = 0.282_{-0.033}^{+0.034}$, and $\sigma = 0.805_{-0.017}^{+0.02}$ \citep[Table 2 of][see also most recently \citet{Li.etal.2025}]{Carlsten.etal.2021}.

Stellar masses of the dwarf satellite galaxies are estimated using the color–$M_\star/L$ relation from \citet{Into.etal.2013} \citep[see, e.g., Section 2.6 in][]{Carlsten.etal.2021}, assuming the candidates are at the distance of their respective hosts. Since the $1\sigma$ uncertainty of the size-mass relation is derived by marginalizing over a posterior distribution and cannot be fully reconstructed, we adopt a flat uncertainty of 0.1 dex on the size. Sources that deviate by more than $5\sigma$ (i.e., $>0.5$ dex) above or $3\sigma$ below the size-mass relation are excluded. We have a wider size range for larger objects to account for the presence of ultra-diffuse galaxies \citep[UDGs, see e.g.][]{van.Dokkum.etal.2015, Greco.etal.2018, Danieli.etal.2019, Danieli.etal.2022, Li.etal.2023a, Li.etal.2023b}.

In crowded sky regions where multiple host galaxies are in close proximity, a candidate satellite galaxy may be selected as a satellite of two or more hosts. To address this issue, we assign each candidate satellite galaxy to its nearest host, thereby eliminating repeated counts. After applying the magnitude, color, and angular size cuts, we initially identify approximately $\sim41,000$ sources across the 393 hosts. This number is reduced to 1191 sources after applying the integrated $p(z)$ cut, and further reduced to 972 sources following the size-mass cut. After resolving duplicates, the sample is reduced to 950 sources.

%---------- SUBSECTION 2.3.5: Deblending
\subsubsection{Deblending} \label{sec:deblending}

Blended objects present another source of contamination in our sample. These include shredded fragments of massive galaxies, blended objects, or instances where multiple parts of the same blended source are identified separately. To ensure the robustness of our sample, we conducted a thorough visual inspection of all 950 sources retained after resolving duplicates. Six authors in the Merian collaboration visually inspected 950 galaxies each, with each galaxy receiving five classifications based on the $gri$ image. During this process, blended objects were identified and excluded.

Our catalog also provides $r$-band and N708-band blendedness values, which are generally greater than 0.1 for blended sources. While applying a blendedness cut of $< 0.1$ could help remove some blended objects, it is not a perfect solution due to the complexities listed above -- approximately 20 blended sources out of the entire $950$ sample would still be missed. Moreover, we find that using the blendedness cut instead of visual inspection would result in the loss of $\sim100$ additional sources in total. However, we have verified that this alternative approach does not change the results of this paper. %Therefore, while the blendedness cut provides a useful automated filter, visual inspection remains a crucial step in ensuring the quality of the sample.

After removing blended objects based on the visual inspection, we obtain a final cleaned sample of 793 sources as satellite candidates. This refined sample is used for all subsequent analyses presented in this work. The full downselection process is summarized in Table~\ref{tab:downselection}.

%---------- SUBSECTION 2.5: Contamination correction
\subsection{Background correction} \label{sec:bkg_sub}

Given the precision of our photometric redshifts ($\sigma_{\Delta z/(1+z)} \approx 0.01$), some candidates are inevitably foreground or background objects. To quantify the level of contamination, we perform an additional background subtraction using 60 empty circular fields, each with a radius of $0.1$ degrees. These fields are carefully selected to avoid proximity to any nearby hosts in terms of projected right ascension and declination. Implementing a local background estimate by drawing an annulus around each host beyond the splashback radius could reflect the local density variations more accurately, but we postpone such an analysis to future work.

We apply the same selection criteria outlined in Section~\ref{sec:sat_select} to these fields. There are $\sim 19,000$ sources left after this cut. Since there are no host galaxies in the background fields, we draw a random spectroscopic redshift (spec-$z$) from the host sample's spec-$z$ distribution \citep[see also e.g.][]{Li.etal.2023b}. This random redshift is used to convert the sources' angular sizes to physical sizes, ensuring that the size--mass cut is applied consistently.

To minimize the impact of cosmic variance on the background density estimates, the 60 random fields are distributed across the entire RA and Dec coverage of the survey. This broad spatial sampling ensures that the background density is representative of the survey region in terms of cosmic variance and variations in depth.

After applying all selection cuts and conducting a visual inspection of candidate sources within the 60 blank fields, we total the number of sources passing these cuts across all fields. Photo-$z$ cuts sources from $\sim 19,000$ to a mere 228, and additional visual inspection eliminates  62 more to reach a final background sample number to 166. To get a distribution of background density, we did a Monte Carlo simulation for the 60 fields with $N = 1000$. For each iteration, we randomly select 60 fields with replacement and count the total background number after all the selection cuts. We then divide the total background number by the combined area of the fields ($60\times \pi\times 0.1^2$ deg$^2$) to get a background density. The median and standard deviation for the background density is $88\pm 10$ sources per square degree. In subsequent analyses, we multiply this density by each host's virial area to estimate the level of contamination for each host. The total background number for all 393 hosts is 342$\pm38$, calculated from multiplying the background density and error by the total host virial area. After background subtraction and accounting for Poisson error of the number of satellite candidates, our sample makes a total of 451 $\pm 47$ satellite candidates. 

When analyzing the radial distribution of satellite galaxies and satellite abundance as a function of host properties, we subtract the estimated contamination from the number of candidate sources. The virial areas of the hosts in our sample range from $0.005\ \mathrm{deg}^2$ to $0.025\ \mathrm{deg}^2$, leading to an expected contamination of approximately $0.44$ to $2.22$ sources per host. This subtraction ensures that our measurements of satellite abundance and spatial distributions reflect true satellite populations, mitigating the effects of background contamination.

%---------- SUBSECTION 2.6: group mass proxy
\subsection{Group Mass Proxy: \( M_{\rm tot} \)} \label{sec:Mtot}

Understanding how satellite galaxy properties and their abundance depend on the halo environment requires a reliable proxy for group halo mass. A well-chosen mass proxy allows for a consistent characterization of group environments, facilitating comparisons across different systems and enabling robust studies of environmental effects on galaxy evolution. A variety of approaches exist in the literature -- including velocity dispersion, richness, and total luminosity -- each with its own advantages and limitations.

Several group mass catalogs are available for application to our host sample. One such catalog is constructed using the updated group finder developed by \citet{Lim.etal.2017} and applied to SDSS DR13. This method builds upon earlier techniques from \citet{Yang.etal.2005, Yang.etal.2007} and \citet{Lu.etal.2016}, offering improved halo assignments across a broad mass range. We also explored alternative catalogs, such as the one based on the full flux-limited SDSS main galaxy sample\footnote{\url{https://www.galaxygroupfinder.net/catalogs}} \citep{Tinker.etal.2020, Tinker.etal.2021b}. However, that catalog has only 265 overlaps with our 393 Merian hosts, making it less suitable for our purposes. In contrast, all 393 hosts in our sample are identified as centrals in the SDSS DR13 group catalog and therefore have assigned halo masses. This catalog estimates halo masses using a ranked abundance-matching technique \citep[e.g.,][]{Yang.etal.2007}, which correlates total luminosity or stellar mass with halo mass under the assumption of a monotonic relation between the two quantities. While this method performs well for large statistical samples, it can be unreliable for individual systems, particularly at the low-mass end where scatter in the luminosity--halo mass relation and variations in group assembly history become significant.

Given these considerations, we instead adopt the total stellar mass of the system, \( M_{\rm tot} \), as our preferred proxy for group halo mass. We define \( M_{\rm tot} \) as the sum of the stellar masses of the host galaxy and all associated satellite candidates in Merian. This choice provides a practical alternative to abundance-matched halo masses: \( M_{\rm tot} \) is straightforward to measure in wide-field imaging surveys and does not require spectroscopic data or dynamical modeling, making it a scalable quantity for our analysis. $M_{\rm tot}$ is also known to correlate more closely with satellite abundance than many other indicators \citep[e.g.,][]{Carlsten.etal.2022, Mao.etal.2024}.

That said, \(M_{\rm tot}\) is also imperfect. Systematic uncertainties in individual stellar mass estimates (e.g., dust attenuation, outshining, metallicity, and IMF assumptions; see Section~\ref{sec:host_select}) propagate directly into \(M_{\rm tot}\), and in this study \(M_{\rm tot}\) is further affected by the statistical background subtraction applied to satellite candidates. Moreover, the correlation between group mass and \(M_{\rm tot}\) is modest (Spearman \(\rho_s = 0.309\); see Appendix~\ref{sec:appendix1}), underscoring that \(M_{\rm tot}\) should be interpreted simply as the summed stellar mass of all satellite candidates associated with each host.

Finally, as introduced in Section~\ref{sec:host_select}, we apply the \texttt{is\_central} criterion from SDSS DR13 to ensure that all selected hosts are centrals. Removing non-central systems mitigates contamination from satellites of larger groups and avoids complications associated with satellites-of-satellites, yielding a cleaner and more physically consistent comparison to the Milky Way, which resides at the center of its own halo.

%% file: sections/3.stats.tex
In the previous section we used Merian medium-band emission line imaging to build a sample of 451 $\pm47$ satellite candidates after background subtraction. Now, we consider satellite radial profile, satellite abundance, and satellite properties in our statistical sample.

%---------- SUBSECTION 3.1: Satellite radial distribution 
\subsection{Satellite radial distribution} \label{sec:sat_rad_dist}

\begin{figure}
    \centering
    \includegraphics[width=\columnwidth]{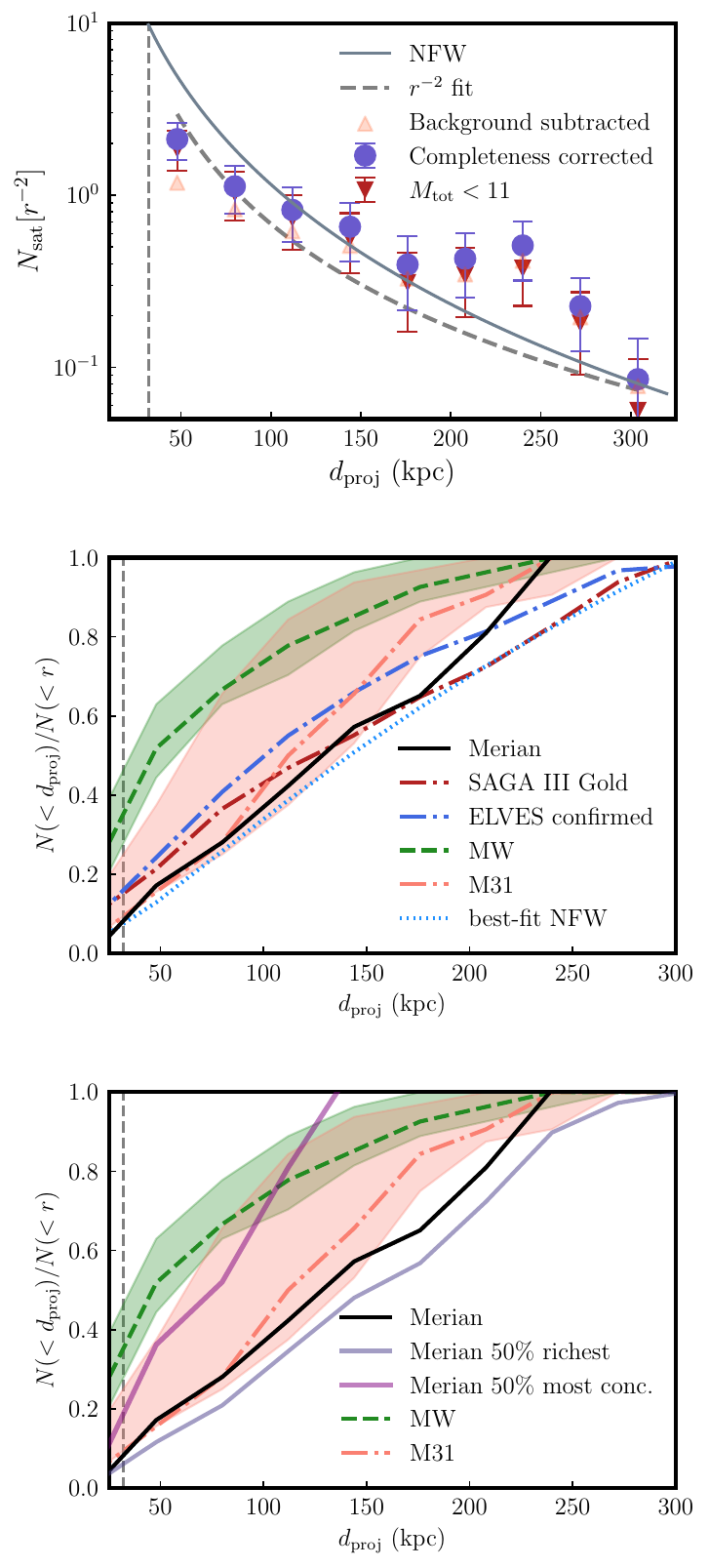}
    \caption{\emph{Top}: Differential satellite radial distribution for the background subtracted sample (orange triangles), completeness corrected sample (purple circles), and systems with $\log[M_{\rm tot}/M_{\odot}] < 11$ (upside-down triangles). We also show the best-fit NFW profile as the solid grey curve with concentration parameter $c_{\rm NFW} = 4.48^{+2.20}_{-1.48}$ and the best-fit $r^{-2}$ profile as dashed grey curve. \emph{Middle}: Average cumulative radial distribution of satellites, for the Merian background subtracted sample (black solid line), SAGA III Gold sample (dashed red line), ELVES confirmed sample (blue dash-dotted line) Milky Way (green dashed line) and M31 (orange dot-dashed line), and best-fit NFW model. \emph{Bottom}: Same cumulative radial profile but now for a subsample of the 50\% most centrally concentrated systems and 50\% richest systems.}
    \label{fig:sat_rad_dist}
\end{figure}

\begin{figure*}
    \centering
    \includegraphics[width=\textwidth]{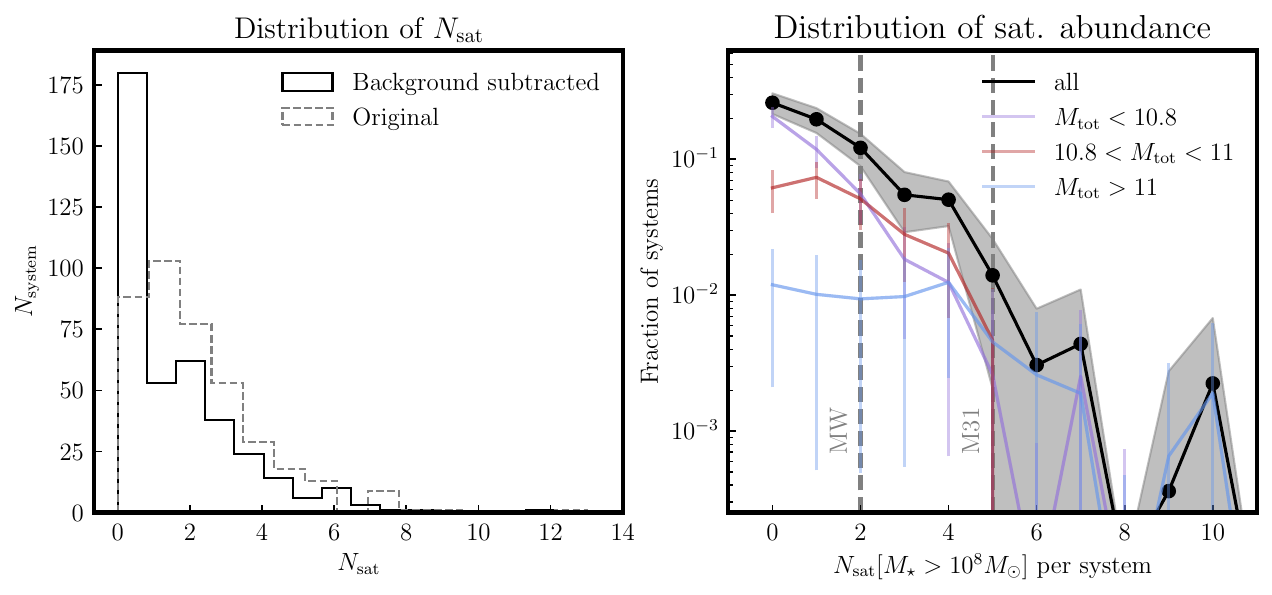}
    \caption{\emph{Left}: the distribution of the number of satellite galaxies $(N_{\rm sat})$ for the 393 hosts in our sample. The dashed line represents the distribution without background subtraction, while the solid line corresponds to the distribution after background subtraction. To ensure a statistically meaningful estimate of $N_{\rm sat}$ in the low-count regime, we adopt a Bayesian approach with a Jeffreys' prior. Specifically, we model the posterior distribution of true satellite counts as a Gamma distribution with shape parameter $k_{\rm obs} - b + 0.5$, where $k_{\rm obs}$ is the number of observed candidates and $b$ is the expected background. We take the posterior mean of this distribution to estimate $N_{\rm sat}$, allowing for a principled treatment of cases where background subtraction yields negative values due to Poisson fluctuations. \emph{Right}: fraction of systems that have satellites with $M_{\star,\rm sat} > 10^{8} M_\odot$ as a function of $N_{\rm sat}$, separated by the system's total stellar mass $M_{\rm tot}$. Errorbars show 16th and 84th percentiles of the distribution calculated from a Monte Carlo simulation with Niter = 1000.  MW and M31 have 2 and 5 such satellites, respectively.
 }
    \label{fig:Nsat_dist}
\end{figure*}

Figure~\ref{fig:sat_rad_dist} presents the radial distribution of satellites through both differential and cumulative profiles. Because our survey targets star-forming, massive satellites, we naturally miss quenched systems, and this incompleteness becomes more severe toward the group center, where quenching is known to be more prevalent \citep[e.g.,][]{Carlsten.etal.2022, Geha.etal.2024}. To account for this effect, we interpolate the quenched fraction, $f_{\rm quenched}$, as a function of $d_{\rm proj}$ from SAGA~III systems \citep[top panel of Figure~5 in][]{Mao.etal.2024} at our radial bins, and divide the background-subtracted radial distribution by $(1-f_{\rm quenched})$ to obtain a completeness-corrected profile.

After applying this correction, we model the satellite distribution using an NFW profile. We use \texttt{colossus}\footnote{\url{https://bdiemer.bitbucket.io/colossus/}} \citep{Diemer.2018} to fit an NFW profile to the completeness-corrected differential distribution, excluding radii with $d_{\rm proj} \lesssim 30 \,\mathrm{kpc}$. The best-fit concentration parameter is $c_{\rm NFW} = 4.48^{+2.20}_{-1.48}$. The smooth decline of the corrected profile aligns closely with the NFW model, suggesting that our satellites broadly trace the underlying dark-matter distribution in their host halos. We additionally show the radial distribution for systems with $\log(M_{\rm tot}/M_\odot) < 11$ to highlight environments most similar to the Milky Way. For comparison, a simple power-law fit with $r^{-2}$ yields a lower $\chi^2 \approx 15$ than our best-fit NFW model ($\chi^2 \approx 40$). Overall, the shape of our differential profile remains broadly consistent with other extragalactic satellite studies, which typically find number-density profiles close to $r^{-1.2}$ \citep[e.g.,][]{Nierenberg.etal.2012}.

The top panel of Figure~\ref{fig:sat_rad_dist} also compares the radial distributions before and after applying completeness and background corrections. These comparisons emphasize that our survey remains incomplete in the innermost region ($d_{\rm proj} \lesssim 30 \,\mathrm{kpc}$), where observational limitations and crowding hinder satellite detection. Many previous studies adopt a similar inner cut of $\sim 0.15$--$0.2\,R_{\rm vir}$ for related reasons \citep[e.g.,][]{vanderBurg.etal.2016, Li.etal.2023b}. Accurately characterizing this innermost region will require deeper imaging and more detailed modeling, which we defer to future work.

The cumulative distribution (bottom panel of Figure~\ref{fig:sat_rad_dist}) allows a broader comparison with SAGA~III Gold, the ELVES confirmed satellite sample, and satellites around the Milky Way and M31, as well as an NFW model with the best-fit values to our data. For the Milky Way and M31, we use the satellite catalog of \citet{McConnachie.2012} and generate projected 2D distributions by mock observing each system along 5000 randomly chosen sight lines. The resulting median curves are plotted as dashed green and dashed–dotted orange lines, respectively, and the shaded regions indicate the 1$\sigma$ variation due to projection. Although these surveys span different limiting satellite masses, $\sim 10^{7.5} M_\odot$ for SAGA~III Gold, $5 \times 10^{5} M_\odot$ for ELVES, $340\,M_\odot$ for the Milky Way, $3 \times 10^{4} M_\odot$ for M31, and $10^{7.1} M_\odot$ for Merian, we find that the cumulative profiles for ELVES and SAGA~III remain largely unchanged when restricting to satellites with $M_\star > 10^{8}\,M_\odot$. Thus, for cross-survey comparisons, we do not impose a uniform lower stellar mass cut. By contrast, applying such a cut to the Milky Way or M31 would leave too few satellites for meaningful analysis, so we include all satellites from \citet{McConnachie.2012}.

The background-subtracted cumulative profile for our Merian sample (solid black line) increases smoothly with radius. When compared to Local Group systems, it reveals notable differences. As observed in both SAGA and ELVES, the Milky Way’s satellites are significantly more centrally concentrated than those in most extragalactic systems, including our own \citep{Carlsten.etal.2020, Mao.etal.2024}. This discrepancy may reflect the presence of the LMC or selection effects that influence satellite identification in external surveys. In contrast, M31 shows a much less centrally concentrated distribution, possibly reflecting different accretion histories between the two Local Group hosts.

To examine variations within our sample, we also analyze two subsamples: the 50\% richest systems, defined by the largest background-subtracted $N_{\rm sat}$, and the 50\% most centrally concentrated systems, defined following \citet{Mao.etal.2024} as those with the lowest median projected satellite distance. The richest systems display a radial profile similar to the overall Merian median, suggesting that richness alone does not strongly reshape the distribution. In contrast, the most centrally concentrated systems show a clear enhancement of satellites at small radii, more closely resembling the Milky Way’s concentrated profile.

Finally, we investigate whether other host properties correlate with satellite radial structure. We explore host color, stellar mass, and several additional indicators of halo concentration but find little variation in the satellite distributions across these properties. This behavior is consistent with the findings of \citet{Wu.etal.2022}, who showed that satellite radial profiles are relatively insensitive to host mass, color, morphology, and the magnitude of the brightest satellite. Together, these results suggest that while a subset of systems can exhibit enhanced central concentrations, the majority of Milky Way analogs, including those in the Merian sample, display more extended satellite distributions that align with expectations from $\Lambda$CDM halo profiles.

%---------- SUBSECTION 3.2: Satellite abundance
\subsection{Satellite Abundance} \label{sec:sat_abundance}

This section explores how the number and distribution of satellite galaxies are shaped by their host properties and environments. By analyzing satellite abundance across a diverse sample of host galaxies, we aim to uncover key insights into the hierarchical structure of the universe and the dynamics governing satellite-host interactions.

\begin{figure*}
    \centering
    \includegraphics[width=\textwidth]{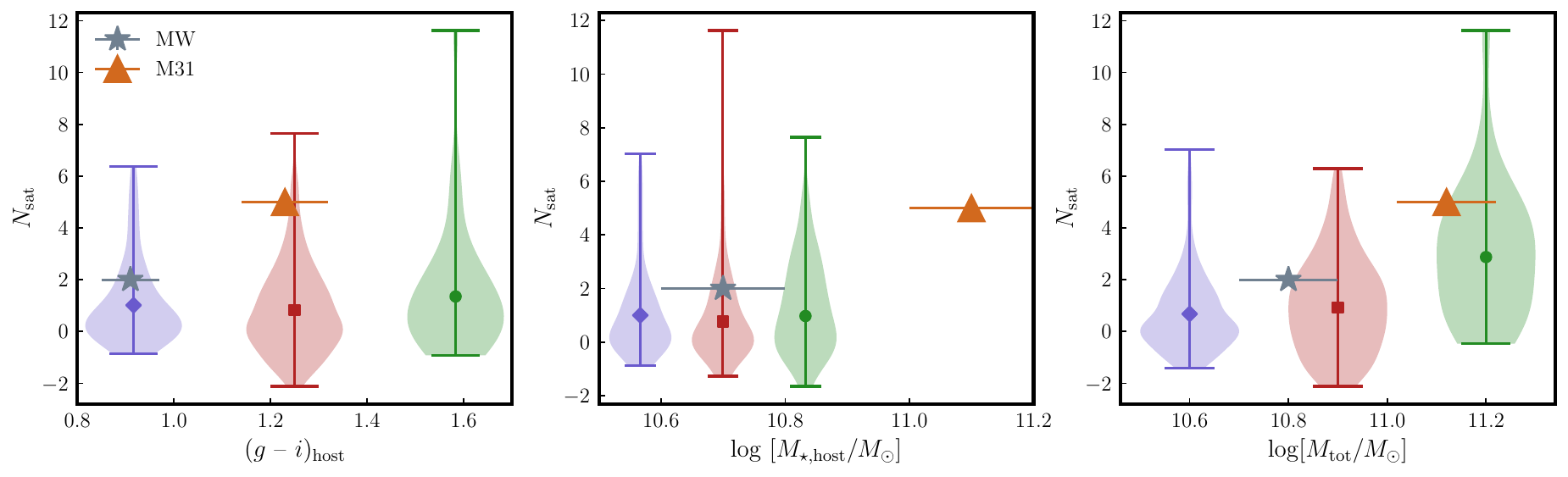}
    \caption{Satellite abundance $(N_{\rm sat})$ as a function of three host properties: host $g-i$ color (left), host stellar mass (middle), and system total stellar mass (right). We bin the host property into three bins. The median $N_{\rm sat}$ in each bin is shown as squares with the minimum and maximum $N_{\rm sat}$ as horizontal bars. Satellite abundance is not strongly correlated with  host color and host stellar mass, but is strongly correlated with the system's total stellar mass $M_{\rm tot}$. We also mark the properties of Milky Way and M31 as grey star and orange triangles, respectively. Milky Way's $g-i$ color is from \citet{Licquia.etal.2015}, stellar mass is from \citet{Licquia.Newman.2015}, and total stellar mass is primarily counting the stellar mass of LMC and SMC from \citet{McConnachie.2012}. M31's $g-i$ color is from \citet{Tempel.etal.2011}, stellar mass is from \citet{Tamm.etal.2012}, and total stellar mass is from \citet{McConnachie.2012}.}
    \label{fig:Nsat_host_prop}
\end{figure*}

Figure~\ref{fig:Nsat_dist} introduces this concept by examining how satellite abundance $(N_{\rm sat})$ – the number of satellites a host galaxy possesses – is distributed across our sample of 393 host galaxies. Since we target massive, LMC/SMC-like, blue satellites, our sample of satellites is roughly complete for $\log[M_{\star, \rm sat}/M_\odot] \gtrsim 8$. This number is calculated by taking the limiting magnitude of our sample (23rd magnitude) and the reddest $g-r$ color (average of the 84th and 95th percentiles), and using the conversion in Eqn~\ref{eqn:mass_color}. 

To estimate the background-subtracted satellite abundance $N_{\rm sat}$ while accounting for Poisson noise in the low-count regime, we adopt a Bayesian approach using a Jeffreys' prior. For each system, we model the posterior distribution of the true satellite counts $\lambda$ using a Gamma distribution with shape parameter $k_{\rm obs} - b + 0.5$, where $k_{\rm obs}$ is the observed number of candidate satellites and $b$ is the expected background count (see more in Appendix~\ref{sec:appendixb}). This posterior arises from a Poisson likelihood combined with a Jeffreys' prior, $P(\lambda) \propto \lambda^{-1/2}$. The background-subtracted satellite abundance is then estimated as the posterior mean of this distribution:
\[
N_{\rm sat} = \mathbb{E}[\lambda] = k_{\rm obs} - b + 0.5.
\]
This formulation provides a smooth and non-pathological estimate of $N_{\rm sat}$ even when $k_{\rm obs}$ is small or zero, while naturally incorporating statistical uncertainty in the background subtraction. This is shown as the solid black line in the left panel of Figure~\ref{fig:Nsat_dist}.

We observe a broad range of massive satellite populations in the Merian hosts -- while some hosts, like the Milky Way, host two massive, star-forming satellites, others are either devoid of or surrounded by a rich entourage of smaller galaxies. The right panel shows that $\sim 80\%$ of our systems have 0 to 3 satellites, with the majority falling toward the lower end of this range. The mean satellite abundance is $0.96\pm1.46$  in our sample. We found that $51\pm 5\%$ systems have no satellites, $19\pm4\%$ have 1 satellite, $13\pm4\%$ have 2 satellites like the Milky Way, and $17\pm4\%$ have 3+ satellites. The error bars come from a combination of Poisson statistics and background estimate uncertainties. We did a Monte Carlo simulation with $N_\mathrm{iter} = 1000$ to calculate the mean and standard deviation of this fraction. For each iteration, we randomly draw 393 systems with replacements and calculate their corresponding background counts by drawing from the background density distribution calculated in Section~\ref{sec:bkg_sub}. We say the system has 2 satellites if after background subtraction the number of satellites fall between 1 and 3, and similarly for $N_{\rm sat} = 3$ and higher. This scarcity of systems with high satellite counts underscores the rarity of massive and highly populated halos in a Milky Way-like environment.

The right panel of Figure~\ref{fig:Nsat_dist} illustrates how satellite populations vary depending on the total stellar mass of their host systems (\(M_{\rm tot}\)) as defined in Section~\ref{sec:Mtot}. By dividing the data into three mass bins, we can observe how galaxies of different scales host their satellite companions. To compute the errorbars, we similarly did a Monte Carlo simulation for $N_\mathrm{iter} = 1000$. For the least massive systems (\(\log[M_{\rm tot}/M_\odot] < 10.8\)), most hosts retain only one or two satellites, and the fraction of systems with $N_{\rm sat} > 3$ is $26\pm7\%$. This mass range is reflective of systems like the Milky Way \citep[see e.g.,][]{Licquia.Newman.2015}, where only two massive satellites (\(\log[M_{\star, \rm sat}/M_\odot] \gtrsim 8\,\)) are present. Moving to intermediate-mass systems (\(10.8 < \log[M_{\rm tot}/M_\odot] < 11\)), the fraction of systems with low \(N_{\rm sat}\) remains significant at around \(10\%\), but the occurrence of richer systems with \(N_{\rm sat} > 3\) increases compared to their lower-mass counterparts. Finally, in the highest mass bin (\(\log[M_{\rm tot}/M_\odot] > 11\)), we see systems with richer satellite populations. These systems, exemplified by M31, can host five or more satellites, with M31 itself sitting at the upper end of the distribution. 

Overall, the trend is clear: more massive hosts tend to have the capacity to host a larger number of satellites at a fixed satellite stellar mass limit, with a $\log[M_{\rm tot}/M_\odot] = 10.6$ host having on average 0.7 $\pm$ 0.8 satellites and a $\log[M_{\rm tot}/M_\odot] = 11.2$ host having on average 2.8 $\pm$ 1.5 satellites. However, even among the most massive systems, the number of satellite-rich hosts declines steeply. This decline reflects the steep mass function of dark matter halos, where fewer massive halos exist to support large satellite populations. These findings align with theoretical expectations of hierarchical galaxy formation \citep[][]{Wechsler.Tinker.2018, Engler.etal.2021, VanNest.etal.2023} and provide valuable insights into the interplay between host stellar mass and satellite abundance.

To explore further how host properties correlate with satellite abundance, Figure~\ref{fig:Nsat_host_prop} shows the connection between $N_{\rm sat}$ and three key host properties: host color $(g-i)$, host stellar mass $(M_{\star, \rm host})$, and the system's total stellar mass $(M_{\rm tot})$, calculated as the sum of host stellar mass and all the satellites' stellar mass.

The left and middle panels reveal statistically no correlation between host color and host stellar mass and satellite abundance: the median satellite abundance is $1.0\pm0.1, 0.8\pm0.1,$ and $0.9\pm0.1$ for the three host $g-i$ color bins, and  $1.0\pm0.1, 0.8\pm0.1,$ and $1.4\pm0.1$ for the three host stellar mass bins. Previous studies have shown that more massive galaxies reside in more massive dark matter halos, which can host a greater number of satellites \citep[e.g.,][]{ Behroozi.etal.2010, Moster.etal.2013}, and this trend of $N_{\rm sat}$ with host stellar mass extends nicely to LMC-mass hosts \citep{Li.etal.2025}. One possible explanation for this correlation is hierarchical structure formation, in which more massive halos experience a greater number of mergers and accrete more satellites over time, leading to the observed correlation between $M_{\star, \rm host}$ and $N_{\rm sat}$ \citep[e.g.,][]{Fakhouri.etal.2010, Rodriguez-Gomez.etal.2016}. 
Here we do not find strong evidence for the correlation between host stellar mass and satellite abundance, primarily because we have a narrow stellar mass range for our hosts. 

However, this is not the case for satellite abundance and a system’s total stellar mass. The right panel reveals a strong correlation between \( M_{\rm tot} \) and \( N_{\rm sat} \), and this trend remains robust across different mass bin choices. In our sample, this correlation is to some extent expected by construction: \( M_{\rm tot} \) is defined as the sum of the stellar mass of the host and the total stellar mass of its satellites. Given that our hosts span a narrow stellar mass range by design, and the satellite stellar masses span a smaller dynamic range, variations in \( M_{\rm tot} \) are primarily driven by differences in the number and total mass of satellites. As a result, a correlation between \( M_{\rm tot} \) and \( N_{\rm sat} \) naturally arises.

Nevertheless, even within the most massive \( M_{\rm tot} \) bin, we observe a wide spread in \( N_{\rm sat} \), ranging from 0 to 12 satellites. This large scatter indicates that while \( M_{\rm tot} \) traces the overall satellite population, there remains significant halo-to-halo variation in satellite abundance at fixed total stellar mass — potentially reflecting underlying stochasticity in satellite accretion or quenching processes. The correlations we find are broadly consistent with previous results from ELVES and other surveys, which also found that systems with higher total luminosity tend to host more satellites \citep{Carlsten.etal.2022, Mao.etal.2024}. Similarly, \citet{Tinker.etal.2021} showed that total satellite luminosity scales approximately linearly with halo mass, reinforcing the idea that \( M_{\rm tot} \) is a useful tracer of the host halo's mass and assembly history.

\begin{figure}
    \centering
    \includegraphics[width=\columnwidth]{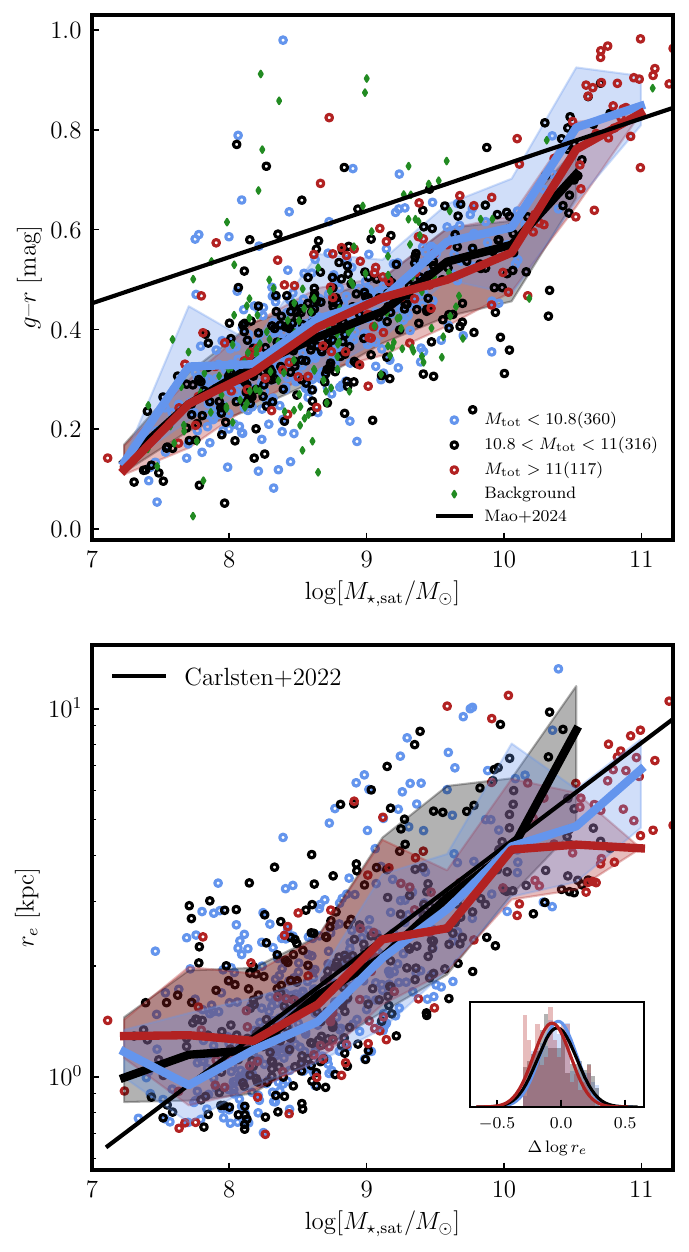}
    \caption{Color (top) and size (bottom) as a function of satellite stellar mass, separated into the same three $M_{\rm tot}$ bins. In the top panel, we also overlay the properties of background galaxies as green diamonds. In both panels, we show the median trend as a function of stellar mass as thick lines for the three $M_{\rm tot}$ bins, and 16th and 84th percentiles as colored bands. The thin diagonal black line in the top panel is the satellite galaxy red/blue division from \citet{Mao.etal.2024}, while the black line in the bottom panel is the late-type dwarf galaxy size--mass relation from \citet{Carlsten.etal.2022}. For the size--mass relation, we put an inset of the distribution of size deviations from the fit ($\Delta \log r_e$) on the lower right corner. For all three subsamples, the distribution is consistent with a Gaussian profile.}
    \label{fig:sat_properties}
\end{figure}

%---------- SUBSECTION 3.3: Satellite properties
\subsection{Satellite properties} \label{sec:sat_properties}

Figure~\ref{fig:sat_properties} examines key physical properties of satellite galaxies – color (\(g-r\)), size (\(r_e\)), and projected distance (\(d_{\rm proj}\)) – as functions of their stellar mass (\(M_{\star, \rm sat}\)). Satellites are categorized into the same three \(M_{\rm tot}\) bins used in our previous analyses, allowing for a direct comparison of trends across different host environments. Together, these panels provide a multidimensional view of how satellite characteristics vary with both their mass and their host’s total stellar mass.

The top panel presents the \(g-r\) color distribution of satellite candidates and our background sample. As expected, given our selection criteria favoring blue, star-forming, LMC/SMC-like satellites, most galaxies in our sample fall bluer than the red/blue division from \citet{Mao.etal.2024}. Above the blue-red satellite color boundary delineated by \citet{Mao.etal.2021}, we see red satellites below \(\log[M_{\star, \rm sat}/M_\odot] = 9\) and above \(\log[M_{\star, \rm sat}/M_\odot] = 10\), but in the range of \(9 < \log[M_{\star, \rm sat}/M_\odot] < 10\) only blue satellites exist. This is qualitatively similar to the satellite quenching time as a function of satellite stellar mass: quenching time is low at both low and high satellite stellar masses and highest in the middle. This non-monolithic trend has been observed in both observations such as ELVES \citep[][]{Greene.etal.2023} and SDSS \citep[][]{Wheeler.etal.2014}, and in simulations \citep[e.g.][]{Wetzel.etal.2013}. At high satellite stellar mass \(\log[M_{\star, \rm sat}/M_\odot] \gtrsim 9.5\), satellite quenching is caused by a combination of gas depletion in the absence of cosmic accretion and internal stellar and black hole feedback \citep[e.g.][]{Wetzel.etal.2015}. Satellite quenching time turns over at low stellar mass \(\log[M_{\star, \rm sat}/M_\odot] \lesssim 9\) due to increased efficiency of environmental process such as ram-pressure stripping \citep[e.g.][]{Simpson.etal.2018, Akins.etal.2021, Pan.etal.2023}. Here in the Merian satellite sample we see the turnover of this quenching time at intermediate satellite stellar mass shown as a lack of red satellites in the mass range \(9 < \log[M_{\star, \rm sat}/M_\odot] < 10\).

The bottom panel shows the effective radius (\(r_e\)) of satellites in kpc units as a function of stellar mass. The observed spread – extending 5\(\sigma\) above and 3\(\sigma\) below the \citet{Carlsten.etal.2022} relation – reflects the size-mass selection cut imposed in Section~\ref{sec:size_mass_cut}. The punchline is that the median size--mass relation and spread is very similar across different environments \citep[see also e.g.][]{Carlsten.etal.2020}. We histogram the residuals, defined as $\Delta \log r_e = \log r_{e,\rm obs} - \log r_{e,\rm fit}$, and find that their distribution is well described by a Gaussian for all three $M_{\rm tot}$ bins with very little skewedness, consistent with the log-normal scatter reported by \citet{Carlsten.etal.2022}.

%% file: sections/4.discussion.tex
Satellite galaxies provide crucial insights into hierarchical galaxy formation and the impact of environment on galaxy evolution. The Merian survey presents a significantly larger statistical census of star-forming satellites compared to previous surveys such as SAGA and ELVES, enabling a robust characterization of satellite abundance, quenching, and structure across diverse host environments. This dataset allows us to examine the Milky Way’s satellite system in a cosmological context, particularly the frequency and properties of LMC-SMC analogs, assessing whether the Milky Way’s satellite configuration is typical or an outlier.

\subsection{A New Statistical Census of Star-Forming Satellites} \label{sec:comparison_SAGA_ELVES}

Merian expands on previous statistical satellites of Milky Way analogs studies \citep[e.g.][]{Tanaka.etal.2018, Wang.etal.2021, Nashimoto.etal.2022, Wang.etal.2025} by substantially increasing the number of hosts while maintaining a consistent mass range and accurate distance measurements via photometric redshifts. Merian’s photometric selection approach allows for the identification of satellites independent of spectroscopic completeness, providing an avenue to study low-surface-brightness satellites and diffuse star-forming galaxies that may be missed in previous spectroscopic surveys.

Figure~\ref{fig:frac_sat_compare} compares the fraction of Milky Way analogs that host $M_\star > 10^8 M_\odot$ satellites as a function of $N_{\rm sat}$ for the Merian, SAGA Gold, and ELVES samples. Since Merian only detects star-forming satellites, we apply a flat correction of $f_{\rm quenched} = 30\%$ \citep[see e.g.][]{Geha.etal.2024} to the fraction of Merian systems across all $N_{\rm sat}$ to account for the missing quenched satellites in our sample. Across most of the satellite number counts range, Merian yields results that are broadly consistent with those from SAGA and ELVES, reinforcing the robustness of our satellite abundance measurements. Thanks to the large host sample, we have much smaller error bars compared to ELVES, especially for $N_{\rm sat} \leq 6$. The figure also places the Milky Way and M31 in context, highlighting that while the Milky Way (with 2 massive satellites) is common, systems like M31 (with 5) are less typical, but M31 has a higher halo mass than the Milky Way and our Merian hosts. These results emphasize the statistical power of Merian in characterizing the abundance of massive satellites and assessing the typicality of the Local Group within the broader galaxy population.

Beyond abundance, we also examine the structural and color properties of satellites in Figure~\ref{fig:sat_properties}. We find red satellites only below \(\log[M_{\star, \rm sat}/M_\odot] = 9\) and above 10, with a lack of red satellites at \(9 < \log[M_{\star, \rm sat}/M_\odot] < 10\). This mass dependence aligns with the non-monolithic quenching time trend -- shortest at low and high masses, longest at intermediate mass -- seen in observations \citep{Greene.etal.2023, Wheeler.etal.2014} and simulations \citep{Wetzel.etal.2013}. High-mass quenching is driven by internal feedback \citep{Wetzel.etal.2015}, while low-mass quenching reflects efficient environmental effects \citep{Simpson.etal.2018, Akins.etal.2021, Pan.etal.2023}. 

Compared to ELVES and SAGA, the Merian sample provides a significantly larger number of star-forming satellites above $10^8\,M_\odot$, enabling improved statistical constraints on the properties of these LMC-like systems. Beyond these local-volume surveys, the recent DESI Bright Galaxy Survey (BGS) analysis by \citet{Wang.etal.2025} measured the luminosity and stellar mass functions of photometric satellites around isolated central galaxies across a broad stellar-mass range ($7.1 < \log M_\star/M_\odot < 11.7$), extending to very faint magnitudes ($M_r \approx -7$). While the DESI BGS study finds that the faint-end slope of the satellite stellar mass function steepens toward lower-mass hosts, Merian focuses on the bright, star-forming end of the satellite population, where individual satellite properties and radial distributions can be robustly characterized. Together, these surveys provide complementary perspectives on satellite demographics.

\subsection{The Milky Way in a Cosmological Context: LMC/SMC-Like Systems} \label{sec:LMC_SMC_context}

An outstanding question in galaxy evolution is the frequency of Milky Way analogs that host an LMC-SMC-like satellite pair. The presence of two massive, star-forming satellites around the Milky Way has often been regarded as a peculiar feature within the broader framework of galaxy formation \citep[e.g.][]{Boylan-Kolchin.etal.2012, Gonzalez.etal.2013, Bose.etal.2020, Haslbauer.etal.2024}. Given the hierarchical nature of structure formation in a $\Lambda$CDM universe, characterizing the abundance and properties of such systems provides insight into both the assembly history of Milky Way-like galaxies and the role of satellite interactions in shaping their evolutionary trajectories. By leveraging the statistical power of our dataset, we can place the Milky Way in a broader cosmological context and quantify the frequency of similar satellite configurations.

As shown in Figure~\ref{fig:Nsat_dist}, we find that approximately 20$\pm2$\% of Milky Way-mass systems in our sample host two satellites with stellar masses comparable to the LMC and SMC ($M_\star > 10^{8.5} M_\odot$). This fraction is in broad agreement with previous estimates from both observational and simulation-based studies. Notably, \citet{Tollerud.etal.2011} used a volume-limited spectroscopic sample from the SDSS to investigate the occurrence of LMC-like satellites around isolated $\sim L_*$ galaxies. Their study found that roughly 12\% of Milky Way analogs host an LMC-like companion within a projected distance of 75 kpc, and 42\% within 250 kpc. We found  11\% of Merian Milky Way analogs host an LMC-like companion ($M_\star > 10^{8.5} M_\odot$) within a projected distance of 75 kpc, and 52\% within 250 kpc. These findings are consistent with expectations from $\Lambda$CDM-based abundance matching models, which predict that $\sim40\%$ of $L_*$ galaxies should host a bright satellite within their virialized halos \citep{Tollerud.etal.2011}. Our results reinforce this picture, showing that the Milky Way’s satellite configuration is not a significant outlier but rather falls within the expected statistical distribution of satellite populations in comparable host environments.

The right panel of Figure~\ref{fig:Nsat_dist} further explores this trend by binning host systems according to their total stellar mass. We find that the fraction of systems with multiple LMC-SMC-like satellites increases with $M_{\rm tot}$, indicating that more massive halos are more likely to host such satellite pairs. This trend aligns with hierarchical galaxy formation, where more massive hosts experience a higher frequency of major mergers and accrete more massive satellites over time \citep{Wechsler.Tinker.2018, Engler.etal.2021, VanNest.etal.2023}. 

Figure~\ref{fig:sat_rad_dist} compares the radial distribution of Merian satellites to those of the Milky Way and M31. We confirm previous findings from the SAGA and ELVES surveys that the Milky Way's satellites are more centrally concentrated than those in extragalactic systems \citep{Carlsten.etal.2020, Mao.etal.2024}. This central concentration remains an outlier even when we examine the radial profiles of the 50\% richest Merian hosts and the 50\% most massive hosts by $M_{\star,\mathrm{host}}$, as these subsets exhibit similar distributions to the full Merian sample. For context, a simple $r^{-2}$ power-law provides a better fit to the Merian radial profile ($\chi^2 \approx 15$) than our best-fit NFW model ($\chi^2 \approx 40$), and the overall shape is broadly consistent with other extragalactic satellite studies that typically find slopes near $r^{-1.2}$ \citep[e.g.,][]{Nierenberg.etal.2012}. When we examine the 50\% most centrally concentrated systems, defined based on the median projected distances of satellites to their respective hosts, we do see an enhanced central concentration but not as strong as the MW. For the cumulative radial distribution comparison, we do not have statistically robust samples for the Milky Way and M31 using only satellites with $M_\star > 10^8\,M_\odot$, so we included all satellites down to ultra-faint mass to facilitate a meaningful comparison with Merian systems. While a uniform lower mass limit is not applied here, future studies that explore how the radial distribution changes with satellite mass thresholds would be valuable.

Although the origin of the observed discrepancy remains debated, our results add further evidence that the Milky Way's environment may be denser or more dynamically evolved than that of typical Milky Way analogs. One possible explanation is that the Milky Way’s satellites have experienced stronger tidal forces and dynamical processing, potentially due to interactions with the LMC and SMC. Alternatively, observational biases in extragalactic satellite surveys -- such as source crowding and confusion at small projected separations -- may partially contribute to the differences in radial distributions.

Overall, our results suggest that the presence of an LMC-SMC-like satellite pair is not anomalous. However, what remains unique about the Milky Way’s satellite system is the orbital configuration and mass ratio of the LMC and SMC, as well as their ongoing interaction with the Milky Way. High-resolution simulations and observational surveys targeting satellites in different evolutionary stages will be crucial in further elucidating the role of such systems in galaxy evolution. Future studies will benefit from combining spectroscopic data and proper motion measurements to reconstruct the detailed orbital histories of these satellite pairs, shedding light on their role in shaping the Milky Way and similar galaxies.

%% file: sections/5.summary.tex
In this work, we present a new statistical census of bright, star-forming satellite galaxies around Milky Way analogs using the Merian Survey. Our sample of 393 host galaxies, significantly larger than previous surveys such as SAGA and ELVES, enables a comprehensive investigation of satellite abundance, stellar mass functions, structural properties, and radial distributions. Our photo-$z$s of the satellites provide accurate distance measurements that significantly boost sample purity against background contamination. Some main findings are summarized below:

\begin{enumerate}
    \item Across the 393 hosts, we identify 793 candidates down to our approximate completeness limit of $M_{\star,\mathrm{sat}} \gtrsim 10^8\,M_{\odot}$. We estimate $451 \pm47$ background-subtracted star-forming satellites in total. 

    \item Our selection is optimized for \emph{star-forming} satellites, so quenched satellites are largely missed, and our counts represent lower limits to the \emph{total} satellite population. Purity and completeness are each $\sim$75\% at our adopted photo-$z$ threshold, and crowding/host light lead to additional incompleteness in the inner $\lesssim$30 kpc.
    
    \item The radial distribution of the background-subtracted and completeness-corrected Merian satellites roughly follows a NFW profile, especially in the outer part $> 80\,\mathrm{kpc}$. Further corrections are needed in the inner part. A simple power-law fit of $r^{-2}$, on the other hand, gives a smaller $\chi^2$ value than the NFW fit. The cumulative radial distribution shows our average Merian satellites is less centrally concentrated than the Milky Way but structurally similar to the SAGA systems. 
    
    \item While $51\pm 5\%$ systems in our sample host 0 satellite,  $19\pm4\%$ have 1 satellite, $13\pm4\%$ have 2 satellites like the Milky Way, and $17\pm4\%$ have 3+ satellites.
    
    \item Comparing the distribution of satellite counts for massive satellites ($M_{\star} > 10^8\,M_{\odot}$) across Merian, SAGA III Gold, and ELVES, we find that Merian is broadly consistent with both surveys across most of the $N_{\rm sat}$ range. Due to its larger sample of host galaxies, Merian provides significantly improved statistical constraints at the massive satellite end, where SAGA and ELVES are limited by small-number statistics. In addition, the recent DESI BGS analysis extends these comparisons to a larger and more diverse population of hosts, finding broadly consistent trends in satellite abundance across host mass, but probing to much fainter limits than achievable with Merian.

\end{enumerate}

The combination of a large sample size, accurate
photometric distance measurements, and statistical completeness at the high-mass end makes Merian a critical step forward in the study of satellite galaxies. By providing a more comprehensive census of star-forming satellites around Milky Way analogs, this survey will play a key role in constraining models of satellite evolution and environmental effects on galaxy quenching.